\documentclass[]{spie}  
\def\bicep{\textsc{Bicep}}
\def\bicepone{\textsc{Bicep1}}
\def\biceptwo{\textsc{Bicep2}}
\def\bicepthree{\textsc{Bicep3}}
\def\biceptng{\textsc{Bicep} Array}
\def\keckarray{{\it Keck Array}}

\usepackage[]{graphicx}
\usepackage{amsmath,amsfonts,amssymb}
\usepackage{xspace} 
\usepackage[hang,flushmargin]{footmisc}
\usepackage{hyperref}
\usepackage{color}

\title{High-Precision Scanning Water Vapor Radiometers for Cosmic Microwave Background Site Characterization and Comparison}

\author{D. Barkats\supit{a}, R. Bowens-Rubin\supit{a}, W. H. Clay\supit{b}, T. Culp\supit{a}, 
R. Hills\supit{c}, J. M. Kovac\supit{a}, \\
N. A. Larsen\supit{b}, S. Paine\supit{a}, C. D. Sheehy\supit{d}, and A. G. Vieregg\supit{b}
\skiplinehalf
\supit{a}Harvard-Smithsonian Center for Astrophysics, Cambridge, MA 02138, USA; \\
\supit{b}Dept. of Physics, Kavli Institute for Cosmological Physics, University of Chicago, Chicago, IL 60605, USA;\\
\supit{c}Cavendish Laboratory, University of Cambridge, CB3 0HE, UK;\\
\supit{d}Physics Department, Brookhaven National Laboratory, Upton, NY 11973, USA\\
}

\authorinfo{Corresponding authors: D. Barkats, 60 Garden Street, MS 42, Cambridge, MA 02138 
USA (dbarkats@cfa.harvard.edu); A. G. Vieregg, 5640 S. Ellis Ave, ERC 429, Chicago, IL 60605 USA (avieregg@kicp.uchicago.edu).}

\begin{document}
  \maketitle

\begin{abstract}

\end{abstract}

The compelling  science case for the observation of B-mode polarization in the cosmic microwave background (CMB) is driving the CMB community to expand the observed sky fraction, either by extending survey sizes or by deploying receivers to potential new northern sites. For ground-based CMB instruments, poorly-mixed atmospheric water vapor constitutes the primary source of short-term sky noise. This results in short-timescale brightness fluctuations, which must be rejected by some form of modulation. To maximize the sensitivity of ground-based CMB observations, it is useful to understand the effects of atmospheric water vapor over timescales and angular scales relevant for CMB polarization measurements. To this end, we have undertaken a campaign to perform a coordinated characterization of current and potential future observing sites using scanning 183~GHz water vapor radiometers (WVRs). So far, we have deployed two identical WVR units; one at the South Pole, Antarctica, and the other at Summit Station, Greenland. The former site has a long heritage of ground-based CMB observations and is the current location  of the \bicep/\keckarray~telescopes as well as the South Pole Telescope. The latter site, though less well characterized,  is under consideration as a northern-hemisphere location for future CMB receivers. Data collection from this campaign began in January 2016 at South Pole and July 2016 at Summit Station. Data analysis is ongoing to reduce the data to a single spatial and temporal statistic that can be used for one-to-one site comparison.



\section{INTRODUCTION}

The study of B-mode polarization in the cosmic microwave background (CMB) has now reached a critical and scientifically exciting phase (see~[\citenum{2016Kamionkowski}] for a recent review of the theoretical and experimental status of the field). In the past five years, the gravitational lensing B-mode signal has been independently detected directly by four different instruments [\citenum{2017PB, 2017ACT, 2015SPT,2016BK}]. At the same time, measurements of B-mode polarization at large angular scales are becoming more sensitive and covering a wider frequency range, with the goal of distinguishing a primordial gravitational wave signal from foreground sources.  Measuring the very weak B-mode CMB polarization signal is extremely challenging, and requires an experimental methodology that can reach the necessary sensitivities while simultaneously controlling systematic errors from the instrument to the stringent levels required for the measurement. Success on both of these fronts is intimately tied to the quality of the atmosphere above the site. 

In the millimeter and sub-millimeter atmospheric windows, water vapor is the primary cause of atmospheric opacity. To minimize excess noise caused by an opaque atmosphere, ground-based CMB telescopes observe in discrete spectral windows, which are bracketed by strong oxygen and water vapor emission/absorption lines. Even within these windows, the water vapor continuum absorbs incoming signal and emits thermally, contributing to noise and reducing the sensitivity of CMB observations. Furthermore, unlike the ``dry'' components of the atmosphere, water vapor is in general poorly mixed, with concentrations that vary dramatically in space and time. CMB telescopes typically rely on fast scanning of the sky to map the relevant spatial scales of the CMB signal onto temporal scales at which their detectors are stable. The time- and space-variable precipitable water vapor (PWV) column along the telescope line of sight causes additional brightness fluctuations in single-dish telescopes (and phase noise/decoherence in interferometers) and limits the spatial scales detectable without relying on other modulation techniques.

The best sites are therefore high-elevation locations with exceptionally dry air. The most competitive ground-based CMB observations to date have taken place at southern hemisphere sites such as the South Pole, Antarctica (the \bicep/\keckarray~program [\citenum{bkii}], the South Pole Telescope (SPT-3G) [\citenum{spt3g}]) and the Atacama Desert in northern Chile (POLARBEAR[\citenum{polarbear2017}], ACTPol[\citenum{actpol2017}], CLASS[\citenum{class2016}], ABS[\citenum{abs2018}]). At the South Pole, the noise properties in CMB maps are well understood due to years of observations from \bicepone, \biceptwo, \bicepthree, \keckarray, and SPT. However, no instruments of this scale are located in the northern hemisphere, and northern sites remain largely untested. The next-generation CMB-S4 experiment~[\citenum{cmbs4}] plans to deploy receivers at the South Pole and Atacama, and is considering locations in Greenland and possibly Tibet.

   \begin{figure}[ht]
   \begin{center}
   \begin{tabular}{c}
   \includegraphics[height=8cm]{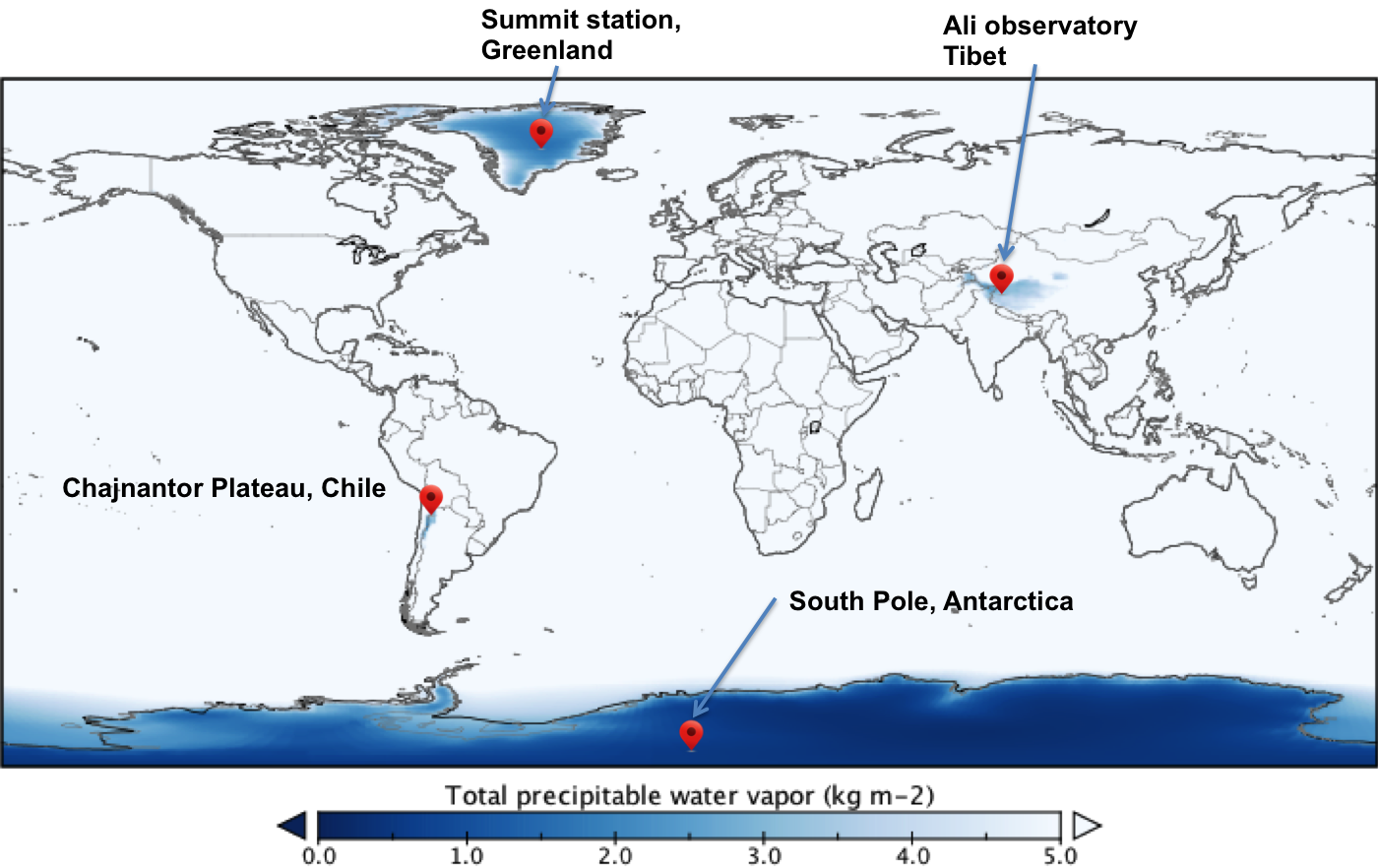}
   \end{tabular}
   \end{center}
   \caption[]
           {Full map of the 10-year average (2006-2016) precipitable water vapor (PWV) estimates from the NASA MERRA-2 reanalysis [\citenum{merra_dataset}]. The four highlighted sites (South Pole in Antarctica, Chajnantor Plateau in the Atacama desert in northern Chile, Summit Station on the polar plateau in Greenland, and the Ali Observatory in Tibet) are all located in the driest zones on Earth.  Note that $1\,kg\,m{^{-2}}$ is equal to 1~mm precipitable water vapor.\label{fig:merra}} 
   \end{figure}


Figure~\ref{fig:merra} shows the 10-year average (2006-2016) Earth map PWV column from the NASA Modern-Era Retrospective Analysis Version 2 (MERRA-2). The MERRA-2 reanalysis assimilates a wide range of measurements from satellites, radiosondes, and surface weather stations into a coupled  model encompassing atmospheric and surface properties, processes, and dynamics [\citenum{merra_dataset, gelaro2017}]. We have highlighted the locations of interest for our study, all of which have very low PWV.

Atmospheric fluctuations can impact observations even in locations with low total PWV.  The effect of atmospheric instability is most obvious in the total-power timestreams of a scanning telescope. The sky variations appear as scan-synchronous fluctuations whose amplitudes grow or shrink as the telescope scans across different parcels of water vapor with more or less PWV. The total power depends strongly on the properties of the atmosphere in a given azimuthal direction, which can vary on a timescale of minutes. Although the emission of the atmosphere is mostly unpolarized (so these fluctuations are highly correlated across detectors and within detector pairs), all polarimeters have only a finite degree of common-mode rejection of unpolarized fluctuations. For any given observing strategy, the amplitude and spectrum of the atmospheric fluctuations dictate the performance and the required level of modulation (observing modulation or software filtering) of the experiment.  


The effects of water vapor on CMB observations can be mitigated by carefully selecting sites with characteristically low total PWV.  Total PWV and its average variation across days and seasons have long been used as a metric for assessing site quality [\citenum{radford_peterson_2016, cortes2016, matsushita2017, sarazin2013, tremblin2011, rose2005}].  Ground-based instruments for recording average PWV are commercially available, and MERRA-2 [\citenum{gelaro2017}] and other reanalyses are now capable of providing accurate site-specific climatologies with temporal resolution on the timescale of hours.

For CMB measurements (as for interferometry), however, what matters most are the spatial and temporal fluctuations of the water vapor on even shorter timescales. There is no substitute for direct measurement of these fluctuations.  The Atacama Large Millimeter/submillimeter Array (ALMA) has led the way in developing sensitive co-pointed water vapor radiometers (WVRs) on each of their 54 12-meter antennas [\citenum{alma_phasecorr_2013}].  The desire to better assess and compare atmospheric fluctuations from site to site [\citenum{layhalverson2000,bussmann2005,sayers2010}] has spawned an interest in developing a small, ultra-precise, scanning-mode WVR that would be appropriate for predicting sky noise for degree-scale CMB experiments.  

We have developed and built such instruments, aiming to make datasets of long-term measurements of atmospheric noise and stability. Our goal is to directly compare leading CMB sites using the same observing strategy with identical instrumentation and analysis. Coordinating measurements across different sites in this way is crucial to directly comparing possible sites for future CMB telescopes. We will also use the understanding of noise levels in CMB maps at the South Pole to translate measurements of sky stability to a prediction for the eventual levels of noise that will remain in CMB maps produced at other sites. Finally, we will use this comparison of sites to comment on the suitability of potential northern hemisphere locations (Summit Station, Greenland and Ali, Tibet) as a complement to the current leading sites (South Pole, Antarctica and Atacama Desert, Chile) for ground-based CMB observation.

\section{INSTRUMENT AND OBSERVING STRATEGY}
Our instrumental design is based on the WVRs commissioned for the ALMA observatory. In addition, we have packaged the WVR unit inside a custom environmental enclosure, along with custom scanning optics in both azimuth and elevation, a temperature control system, and a data acquisition system. See Figure~\ref{fig:schematic} for a schematic of the full system. We have so far built and deployed two of these stand-alone weather-hardened units.

\subsection{Omnisys 183-GHz Water Vapor Radiometer}
\begin{figure}[ht]
   \begin{center}
   \begin{tabular}{c}
   \includegraphics[height=5.5cm]{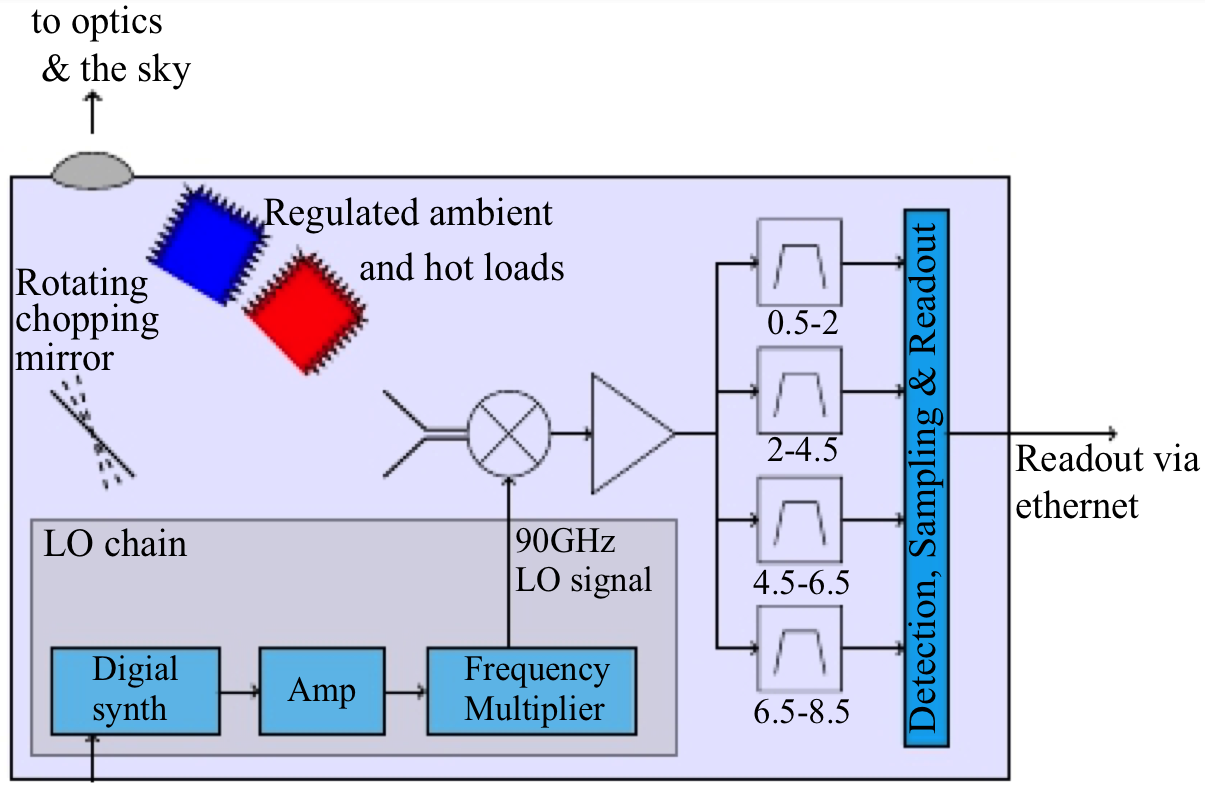}
   \includegraphics[height=6.5cm]{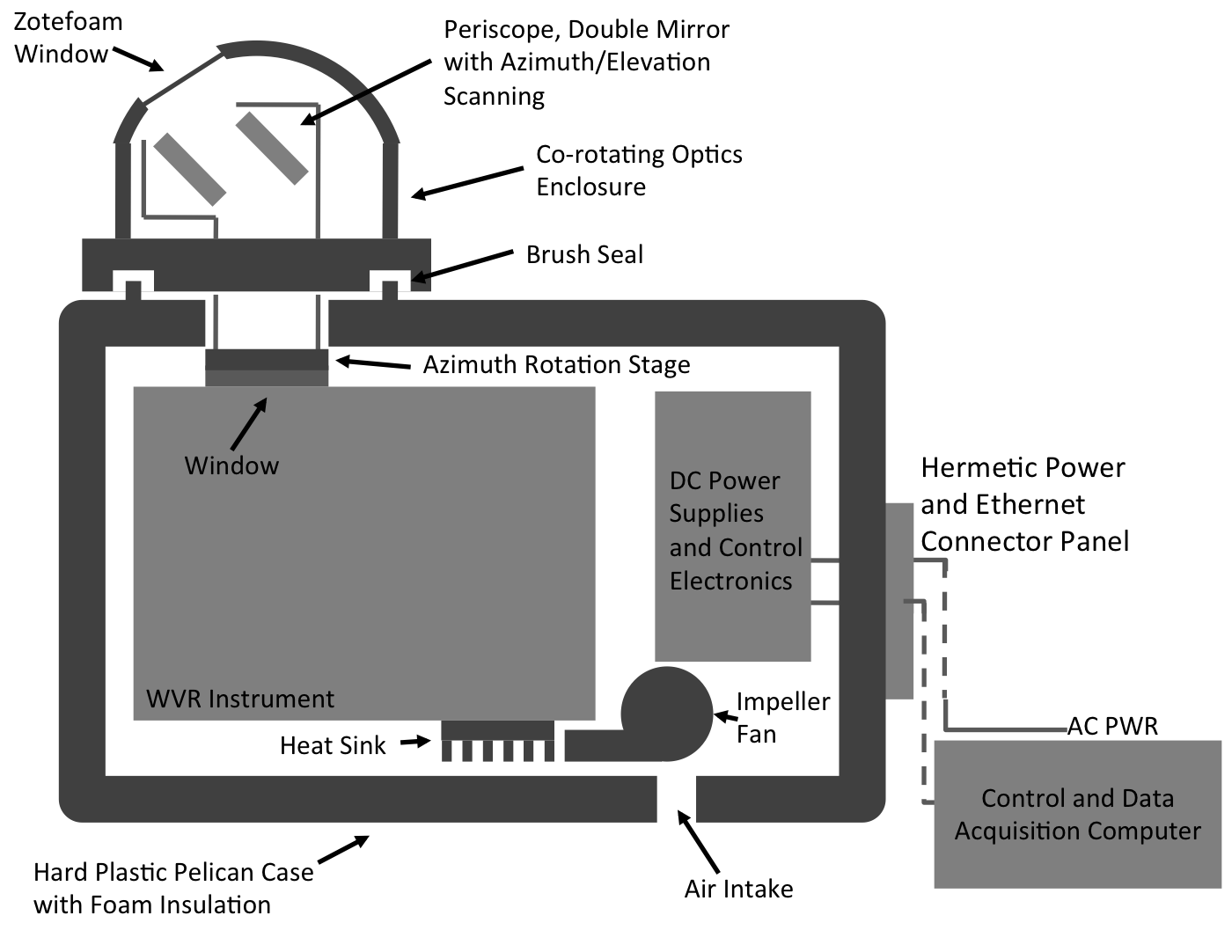}
   \end{tabular}
   \end{center}
   \caption{\textbf{Left:} Schematic of the Omnisys radiometer base unit.  The radiometer is a single-channel, Dicke-switched, warm 183~GHz radiometer.  The input is coupled alternately to the sky, a regulated hot load, and a regulated ambient load by a mirror mechanically chopped at 5~Hz. The input signal is then mixed down from 183 GHz with a double sideband Schottky sub-harmonic mixer, amplified, and split into four frequency bands (0.5-2 GHz, 2-4.5 GHz, 4.5-6.5 GHz, and 6.5-8.5 GHz). These four passbands correspond to the upper and lower sidebands of the 183 GHz water line shown in Figure~\ref{fig:bandpass}. \textbf{Right:} Schematic of the WVR system, including the environmental enclosure, power supplies, scanning optics, and readout system.
     \label{fig:schematic}} 
   \end{figure}

For the base unit, we partnered with Omnisys Instruments\footnote{http://www.omnisys.se/}, leveraging previous R\&D efforts that resulted in the approximately 60 WVR units built for ALMA [\citenum{omnisys_webpage, omnysis_paper}]. A schematic of the instrument is shown in Figure~\ref{fig:schematic}.

\begin{figure}[ht]
   \begin{center}
   \begin{tabular}{c}
   \includegraphics[height=5.5cm,trim=0mm 0.5mm 0mm 0mm,clip]{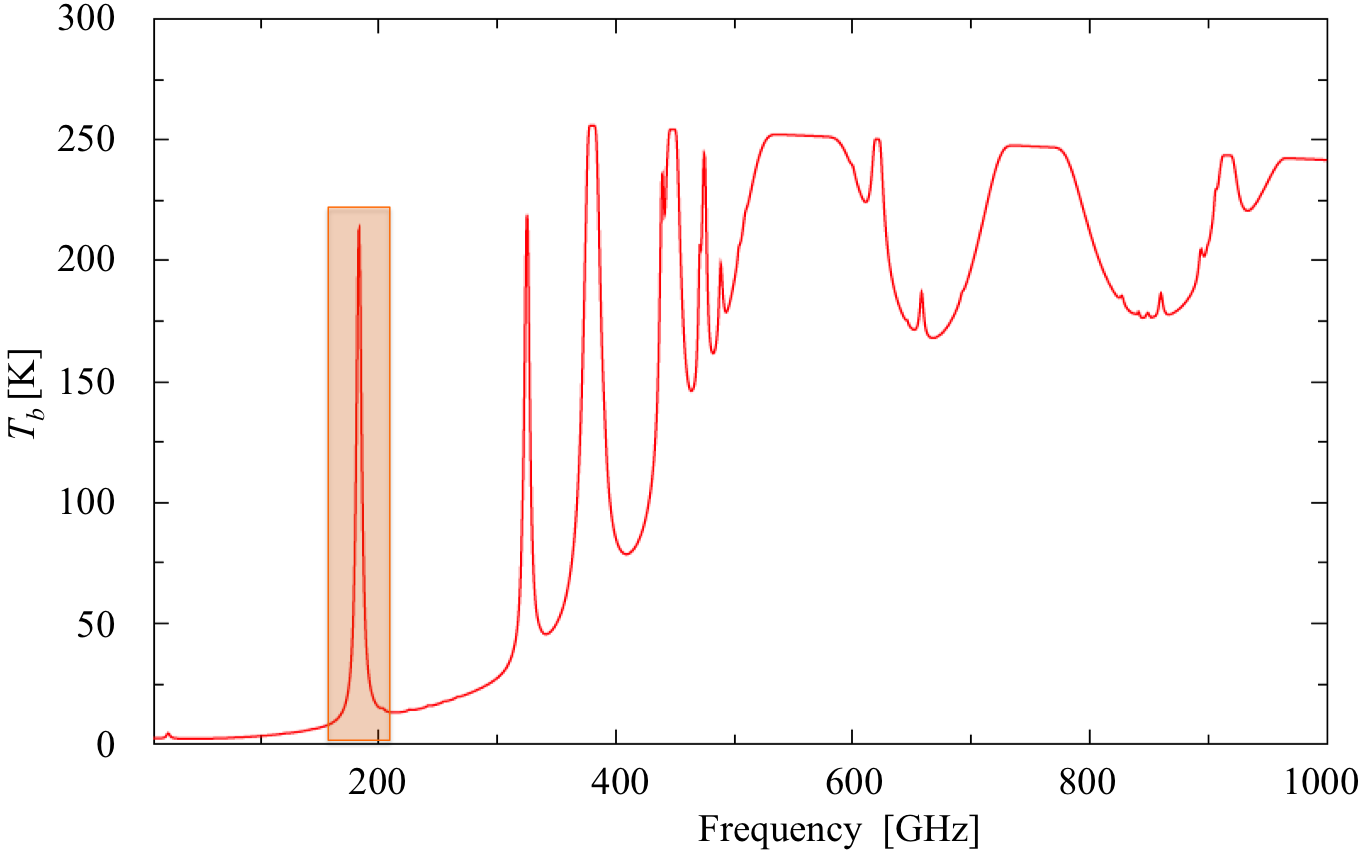}
   \includegraphics[height=5.5cm,trim=0mm 0.5mm 0mm 0mm,clip]{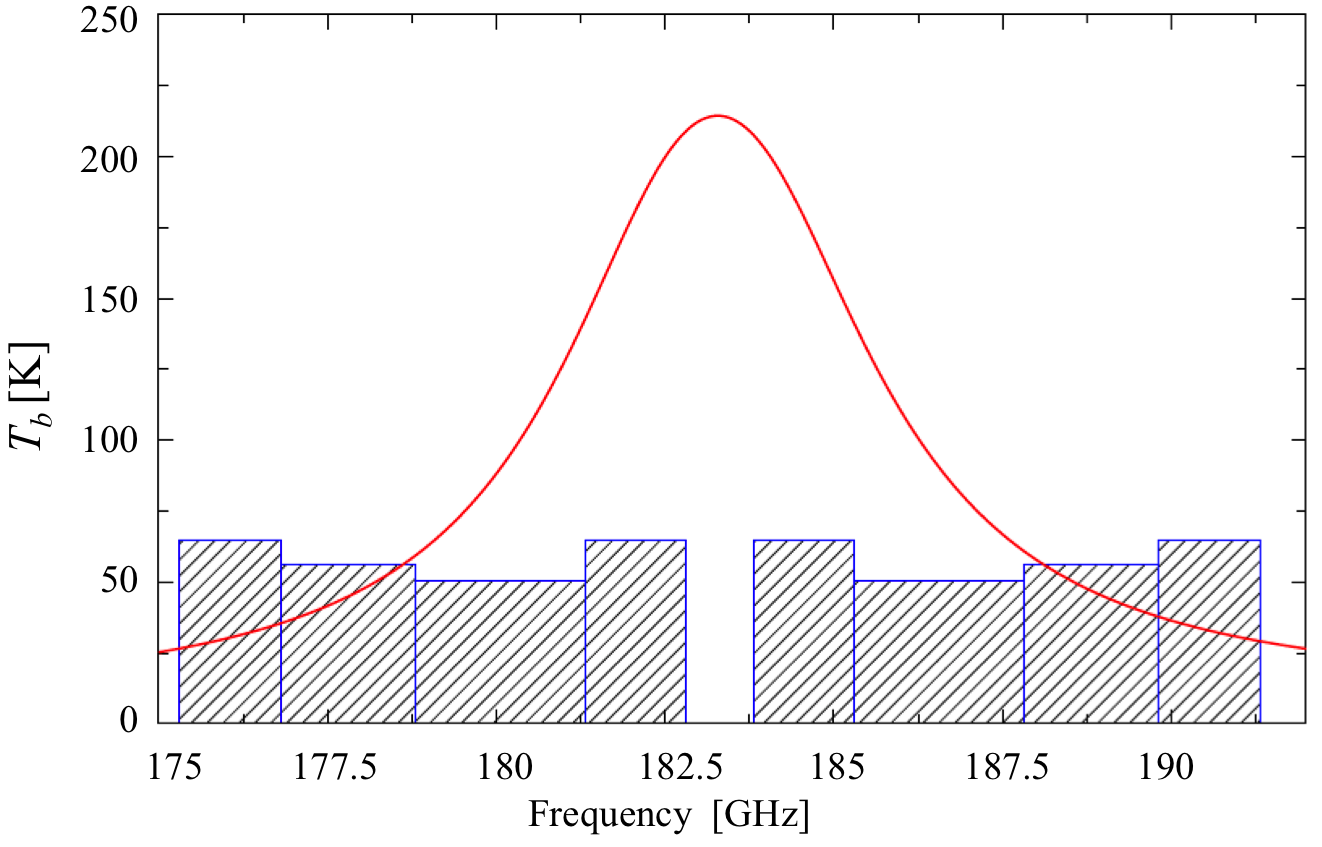}
   \end{tabular}
   \end{center}
   \caption{\textbf{Left:} Atmospheric brightness temperature (defined as $T_{RJ}(\nu) = \frac{c^2}{2k\nu^2}I(\nu)$, where $I(\nu)$ is the spectral radiance) as a function of frequency estimated using \textit{am}~[\citenum{paine_scott_2018}] for a typical atmospheric profile on the Chajnantor plateau in Chile (PWV=1mm). The 183~GHz water vapor emission line is highlighted.
     \textbf{Right}: Close-up of the 183~GHz line and  frequency coverage of the four sidebands of the WVR instrument. The center frequency and width of these bands were chosen to optimize sensitivity to water vapor in a wide range of PWV conditions [\citenum{alma352, alma568, alma495}]. The upper and lower sidebands are shown because the radiometer is double-sideband.
     \label{fig:bandpass}} 
   \end{figure}

The Omnisys WVR is a warm, double-sideband, Dicke-switching 183~GHz radiometer with four spectral channels that straddle the 183.31~GHz water vapor emission line.  Figure~\ref{fig:bandpass} shows the atmospheric brightness temperature spectrum and the nominal bandpass of the four channels. The radiometer incorporates a hot and ambient load, each regulated to better than 1 mK. Dicke-switching (at 5 Hz) between the hot and ambient internal loads provides long-term stability and real-time calibration/conversion into sky brightness temperature units.  We will then use a stand-alone atmospheric radiative transfer model \textit{am}~[\citenum{paine_scott_2018}] to infer the PWV from the four observed sky brightness temperatures. The measured sensitivity of the ALMA WVR is 1~micron of PWV rms in 1~second in the driest conditions (PWV$<$1~mm)[\citenum{alma_phasecorr_2013}]. From our experience at the South Pole, 1~micron PWV is approximately equivalent to a 5~mK additional noise temperature at 150~GHz, an excellent match to the sensitivity needed to perform a useful CMB site characterization.

The Omnisys WVR was designed to operate in the well-regulated environment of the ALMA receiver cabins. Specified performance requires an operating temperature between 16$^\circ$C and 22$^\circ$C. To maintain its flexibility as a stand-alone instrument, we have enclosed it in a custom regulated environmental enclosure that can maintain this narrow temperature range while deployed in extreme polar conditions (temperatures between -20$^\circ$C and -75$^\circ$C and wind speeds up to 40 knots). See Section 2.3 for details. We communicate with the WVR via TCP/IP protocol, unlike the ALMA WVRs which use CANBus, and we have developed a set of Python libraries to control the unit and acquire data from it.


\subsection{Scanning Optics System}
\begin{figure}[ht]
   \begin{center}
   \begin{tabular}{c}
   \includegraphics[height=6cm]{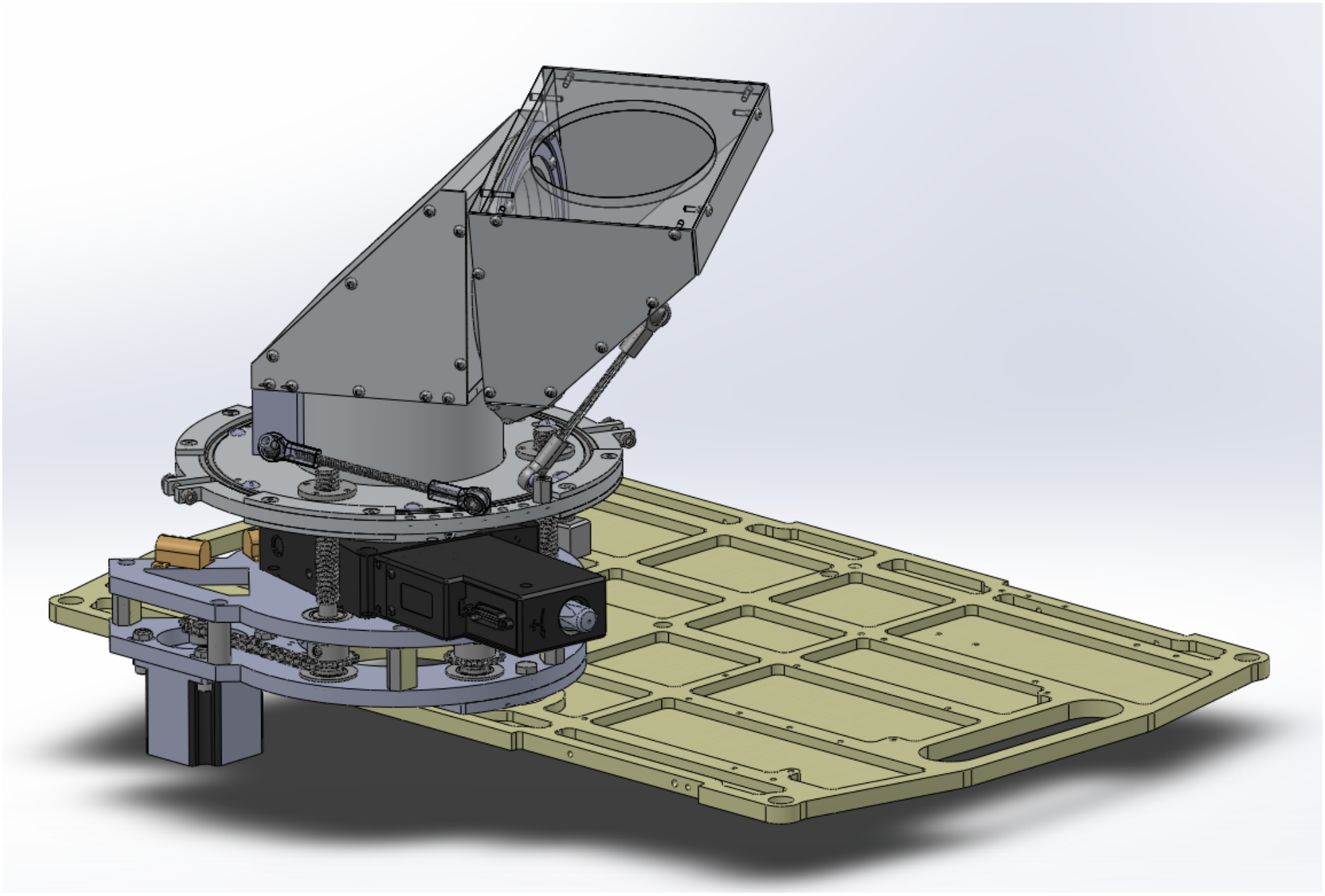}
   \includegraphics[height=6cm]{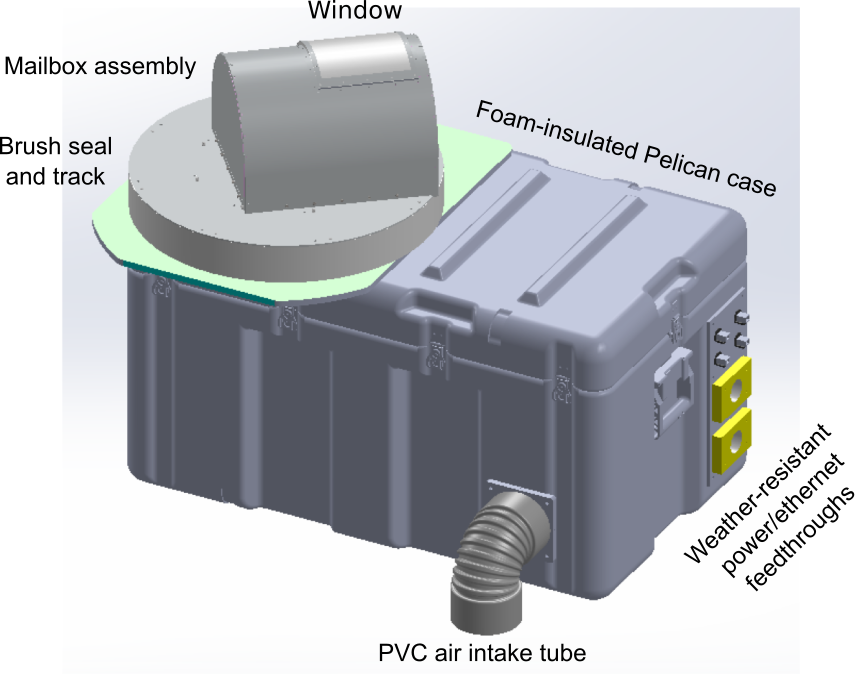}
   \end{tabular}
   \end{center}
   \caption{\textbf{Left:} A CAD model of the scanning optics (the periscope), custom-built for the radiometers.  The opto-mechanical design allows for a
     continuous $360^\circ$ azimuth scanning capability and full elevation scanning capability while eliminating the need for a cable wrap or slip ring. Note that this shows a model of the original commercial rotation stage. A more weather-hardened custom rotation stage was built and deployed in Dec 2017. \textbf{Right:} A CAD model of the environmental enclosure and scanning window, designed to operate in extremely cold and windy environments. 
     \label{fig:optics}} 
   \end{figure}

Our custom-built optics system is designed around two $45^\circ$ flat aluminum mirrors in a periscope configuration. One mirror is fixed on top of the boresight and redirects the beam from the WVR horn to a second mirror that can scan in elevation. Both mirrors are located on a rotation stage that scans in azimuth. This optics system is capable of scanning in elevation between  $10^\circ$ and  $91^\circ$ 
and full continuous $360^\circ$ azimuth scanning.  A schematic of the scanning optics system is shown in Figure~\ref{fig:optics}. This ``periscope'' design combined with the design of our rotation stages allows for azimuth and elevation rotation without a cable wrap or slip ring.

The azimuth rotation stage is a custom-built mechanism using a stepper motor, a belt, bearings, and a static tensioning system. The elevation stage is driven by a separate stepper motor, which synchronously turns three large vertical threaded shafts via a chain.  The shafts are threaded into nuts on the elevation stage, which moves up and down as the stepper motor turns. This stage is connected to the upper portion of the optics tube so that vertical motion of the elevation stage results in elevation rotation of the upper part of the tube.  There is a freely-rotating bearing around the outer edge of the elevation stage to allow for the mapping of vertical motion to rotation of the optics tube.  The upper portion of the optics tube is connected to the rest of the optics system with a bearing.  Both azimuth and elevation stepper motors are controlled using an Arduino\footnote{https://www.arduino.cc/} single board microcontroller and an Arduino Motor Shield. Once an hour, we home the azimuth using an optical transducer, zeroing the azimuth to better than $0.1^\circ$. We use a mechanical limit switch to zero the elevation position prior to any move in elevation. 

\subsection{Environmental Enclosure and Temperature Control System}
The environmental enclosure is designed to allow for operation in extremely cold environments, while maintaining the WVR unit and the moving components of the scanning optics within a 16$^\circ$C to 22$^\circ$C temperature range.  The environmental enclosure is shown in Figure~\ref{fig:optics}. The enclosure itself is a large Pelican\footnote{http://www.pelican.com/} case insulated with two~inches of extruded polystyrene foam. A small air intake keeps the inside of the enclosure under positive pressure. The air flow inside the enclosure is intended to force the warm air out through the top of the enclosure, reducing potential snow and ice accumulation around a dynamic brush seal that allows the periscope to rotate freely with respect to the enclosure. The optics system is entirely contained within the environmental enclosure. As the optics rotate, an insulating foam skirt rotates with it, in contact with a circular brush seal that provides a dynamic seal. The foam skirt and brush seal have a convoluted design to prevent horizontal winds from directly penetrating the enclosure. The rotating part of the environmental enclosure also includes a microwave-transparent window (made of Zotefoam\textsuperscript{\textregistered} HD30) that co-rotates with the optical system.

We have implemented a linear PID temperature control system that delivers between 0 and 230~Watts to the air inside the enclosure. The heater circuit is based on a OPA-541 power amplifier\footnote{http://www.ti.com/lit/ds/symlink/opa541.pdf} that can provide up to 5.5~Amps through eight 1~$\Omega$ power resistors. The resistors are placed on a large heat sink. A high throughput impeller fan forces air through the heat sink to exchange the heat to the air and distribute it around the enclosure. A set of 12 temperature sensors sample the temperature throughout the box and feed a PID temperature control loop on an Arduino inside the enclosure. The temperature control loop functions independently of the data acquisition and communication with the control computer. The power supplies and control electronics for the scanning optics and temperature control systems are also located inside the enclosure, generating additional heat over the baseline. During operations, we have only reached the maximum heating capacity during the coldest times at the South Pole when the temperature dipped below -70$^\circ$C.

\section{OPERATIONS AND DATA ANALYSIS}

\subsection{Deployed Units}
\begin{figure}[ht]
   \begin{center}
   \begin{tabular}{c}
   \includegraphics[height=6cm]{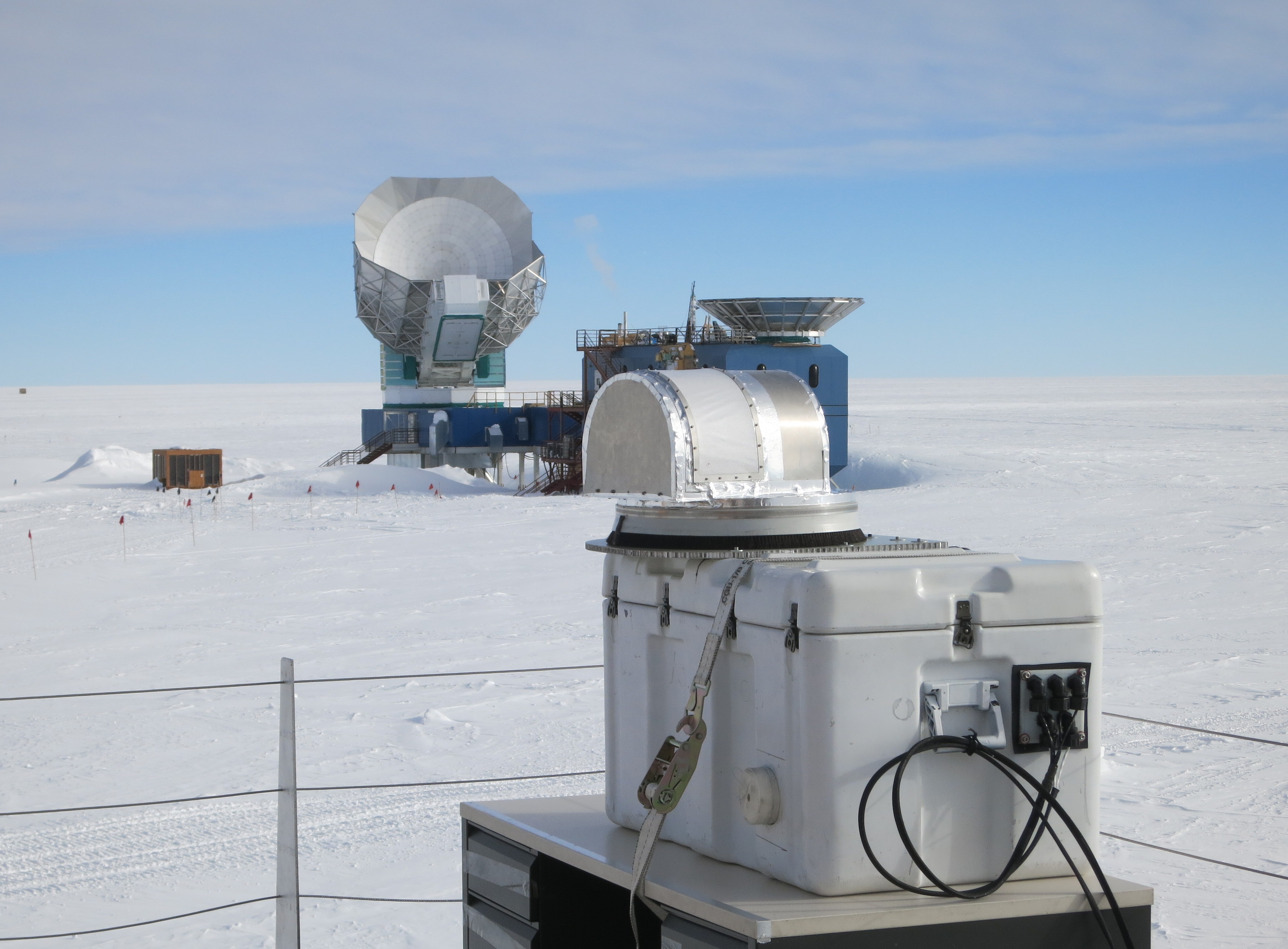}
   \includegraphics[height=6cm]{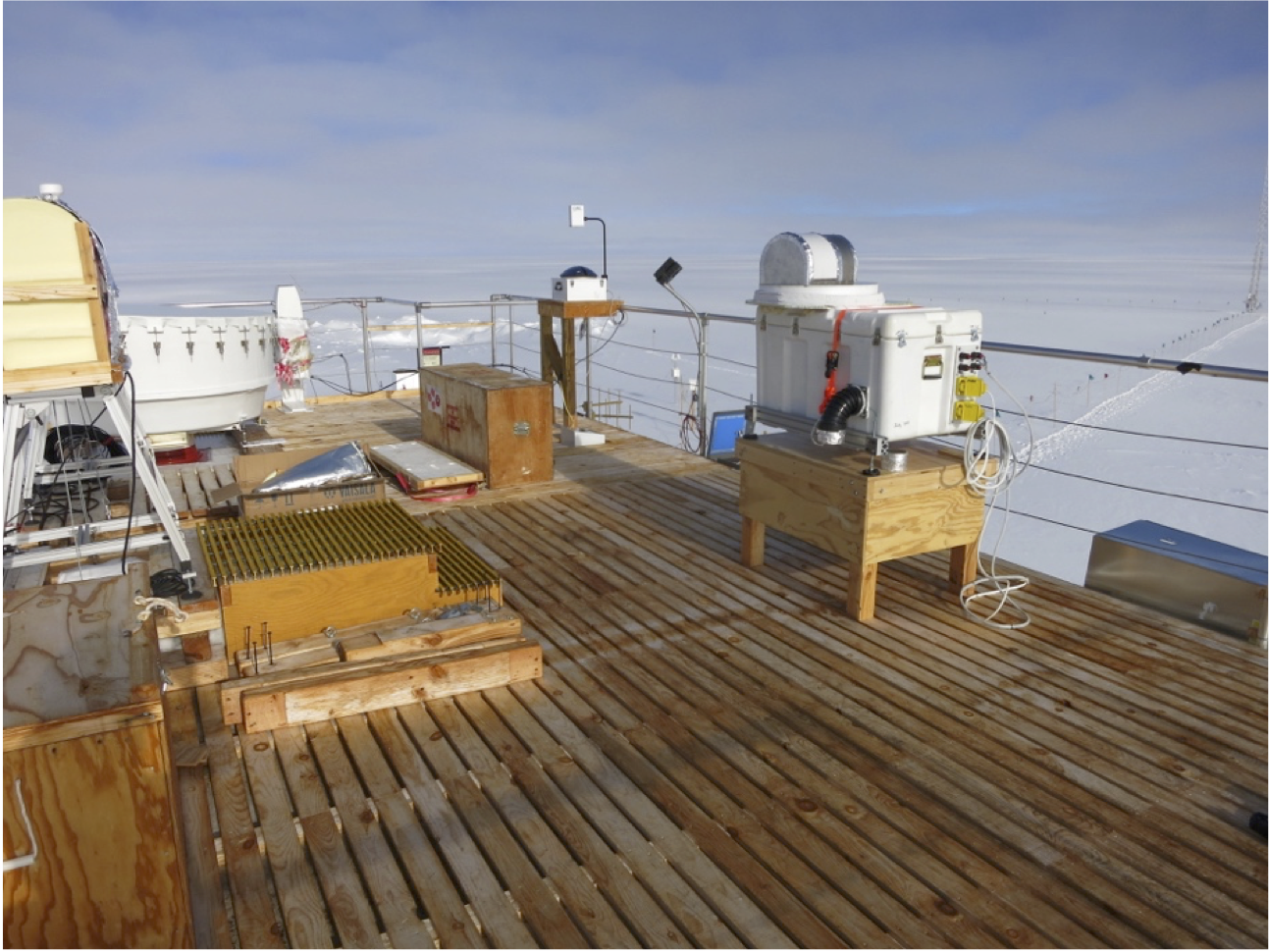}
   \end{tabular}
   \end{center}
   \caption{Two identical WVR units have been deployed.  \textbf{Left:} South Pole station in Antarctica, co-located with
     the Keck Array telescope on the MAPO building (January 2016 -- present).
     \textbf{Right:} Summit Station in Greenland (July 2016 -- April 2017). \label{fig:southPolePic}} 
   \end{figure}

We have so far deployed two WVRs to the field. The first was deployed to the South Pole in January 2016 and is still in operation at that site. The second was deployed to Summit Station, Greenland in June 2016 and operated there until April 2017, when we decommissioned the unit in anticipation of deploying it to an additional site in the future. Photographs of the units deployed at both locations are shown in Figure~\ref{fig:southPolePic}. On both units, the original mechanical design of the azimuth rotation stage failed partway through the winter, preventing further scanning. In January 2018, we upgraded this subsystem on the South Pole unit from a commercial rotating stage to a custom-built more robust rotation stage, which has so far improved the reliability of the azimuth motion.

\subsection{Observing Strategy}
We observe the sky in full, continuous azimuth scans at an elevation of $55^\circ$. The elevation is chosen to match the typical elevation of our co-located CMB instruments. Although feasible, we do not scan in elevation during azimuth scans in order to keep the initial analysis straightforward. The beam FWHM is $\sim8^\circ$. We scan at 2~RPM ($12^{\circ}$/sec) to ``freeze'' atmospheric fluctuations between successive scans. In addition, we perform a  sky dip in elevation every hour to extract the zenith temperature and opacity. Outside of maintenance and downtime due to failures, observations proceed continuously.

\subsection{Ongoing Analysis}
The data analysis closely follows a typical CMB pipeline analysis: from timestreams to maps, from maps to a metric of the spatial and temporal atmospheric fluctuations (akin to power spectra), and from those to statistical measurements of the atmospheric structure over different timescales.

\begin{figure}[ht]
   \begin{center}
   \begin{tabular}{c}
   \includegraphics[height=6cm,trim=0mm 0mm 0mm 0mm,clip]{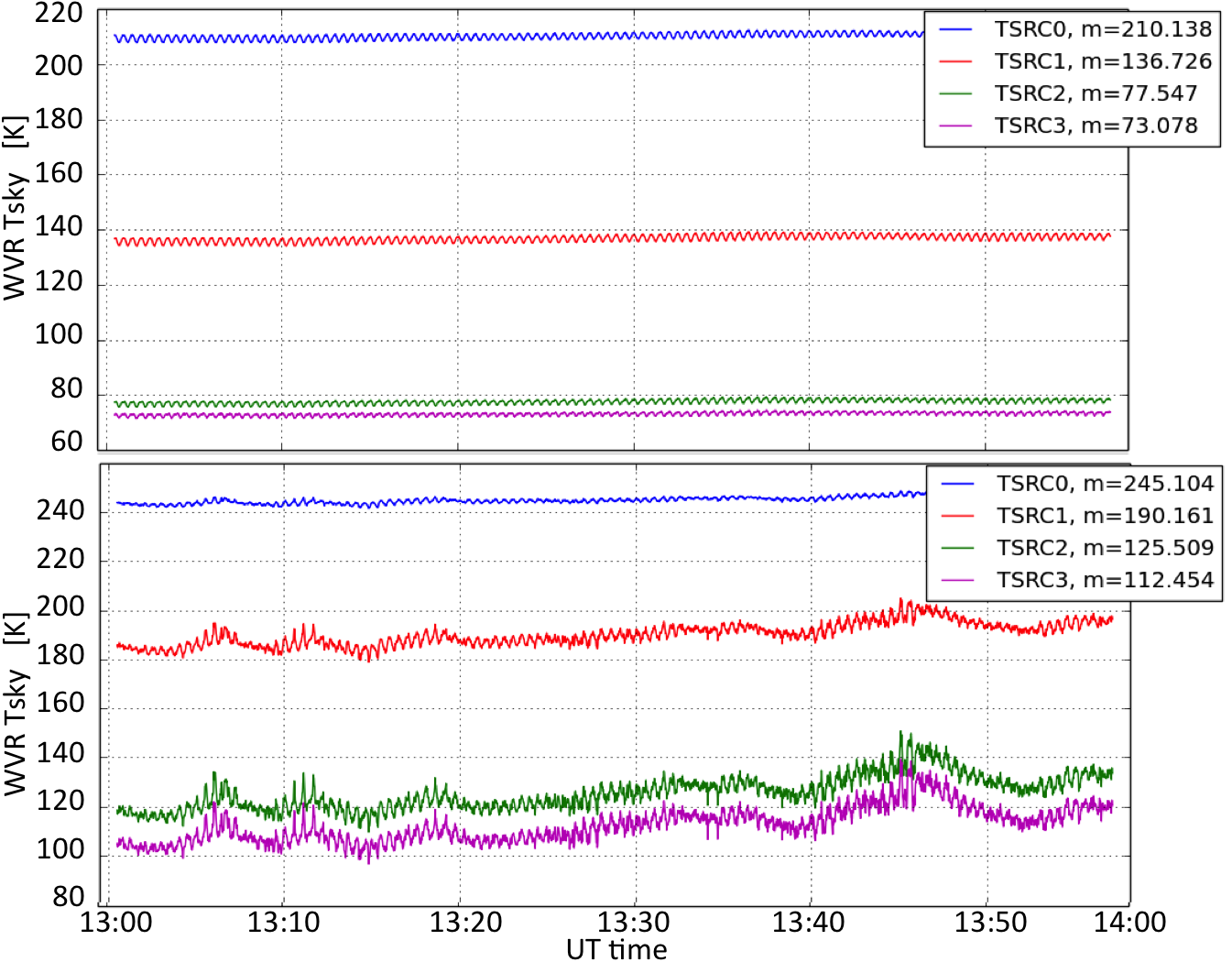}
   \includegraphics[height=6cm]{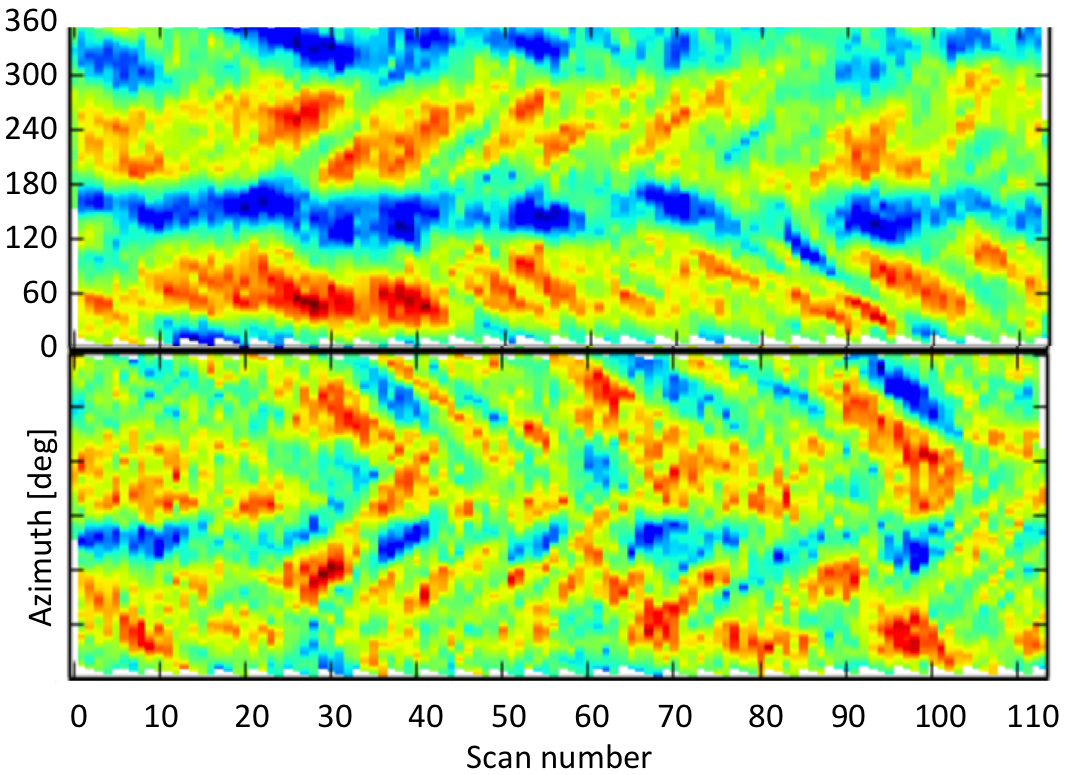}
   \end{tabular}
   \end{center}
   \caption{Initial reduction of one hour of continuous $360^\circ$ azimuth scanning data.  \textbf{Left:} 
     Timeseries of the brightness temperature in Kelvin (Rayleigh-Jeans) for each of the four WVR channels. The small tilt in the azimuth mount causes a scan-synchronous periodic signal, which can be removed in the analysis. The total PWV can be inferred from the relative brightness in each of the four channels. The flat and unchanging timeseries in the top panel denotes a period of very stable and low PWV conditions. A comparatively less-dry and  unstable period is shown in the bottom panel. In addition to the absolute brightness of the four channels being higher on the more humid day, the timeseries show short temporal disturbances corresponding to fluctuations entering and exiting the field of view as the instrument scans. 
     \textbf{Right:} One hour of timestream data for the channel closest to the water line (Channel 0) mapped onto a space-time ``atmogram'' with time on the x-axis and a full azimuth scan on the y-axis.  PWV fluctuations appear as diagonal stripes that come in and out of the field of view. The constant effect of the azimuth mount tilt has been subtracted, though not perfectly, leaving a constant residual 2-theta pattern. The top and bottom panels show one-hour periods separated by 12 hours  ,  when the wind direction has rotated by $180^\circ$, so the orientation of the stripes (fluctuations in the atmosphere) has reversed. \label{fig:tods_maps}} 
   \end{figure}

The initial reduction of one hour of data from the South Pole is shown in Figure~\ref{fig:tods_maps}. On the left, we show the brightness temperature for each of the four WVR channels in a timeseries during a dry and stable period (top) compared to brightness temperature measurements for a humid and unstable period (bottom). On the right, we show one hour of timestream data for the channel closest to the 183~GHz water line (Channel~0) mapped onto a space-time ``atmogram''. PWV fluctuations appear as diagonal stripes that come in and out of the field of view. In the time between the top and bottom panels on the right, the wind changed direction by $180^\circ$. This is evident by the change in the orientation of the diagonal stripes in the ``atmograms''.

Further analysis will involve converting the four measured sky temperatures into a single estimate of the line-of-sight PWV. We will then characterize those fluctuations using parameters of the Kolmogorov-Taylor turbulence model, similar to what was done in ~\citenum{layhalverson2000,bussmann2005,sayers2010}. We can then use those atmospheric power spectrum parameters to statistically compare one site to another, or to compare the fluctuations to those measured by our co-located CMB telescopes. These data can also be used for atmospheric studies for other applications.

\section{DISCUSSION AND FUTURE WORK}
We have developed two stand-alone, weather-hardened, high-precision scanning WVRs that have taken measurements in Greenland and at the South Pole. We plan to extend these measurements to various locations in Atacama and Tibet. These instruments provide a continuous year-round record of water vapor fluctuations to 1~$\mu$m PWV precision, from sub-degree to dipole scales and on timescales $\ge$1~s. In addition, we intend to use these measurements as absolute zenith temperature calibration for co-located CMB experiments. The goal of this program is to collect and publish one-to-one data from different sites that can inform all groups performing or proposing CMB experiments.  It will be the first time there are identical long-term sky noise/stability measurements made to directly compare  leading sites. 

The next-generation, ground-based experiment CMB-S4 is planning to deploy telescopes in both Chile and at the South Pole, and possibly in the Northern Hemisphere, where there are no major current CMB experiments and site quality for CMB observation is largely untested.  Our campaign will help compare the expected performance of telescopes at different sites, including northern sites.

Correlating the distributions of spatial and temporal water vapor fluctuations with the quality of co-measured CMB data at well-characterized sites (such as the South Pole with the \keckarray, \bicepthree, and \biceptng~in the future) gives us a new portable tool to quantify the quality of proposed sites for future large scale CMB observations.  Maintaining this common instrumental and analysis framework is essential to the usefulness of this site characterization data.

\acknowledgments     
We would like to thank the people who supported this project through the winters at the South Pole 
and at Summit Station, including the South Pole Winterovers for \bicepthree~and 
the \keckarray~(Hans Boenish, Grantland Hall, and Robert Schwarz) and 
the winter crew science technicians at Summit Station (Hannah James and Sam Dorsi).  We would also like
to thank Matt Okraszewski, who supported the deployment at Summit Station.  We are indebted to
those at CH2MHill, ASC, and NSF who dedicate their careers to making our science happen
in such remote environments.  This work was funded by the Kavli Institute for Cosmological Physics
at the University of Chicago, the Harvard-Smithsonian Center for Astrophysics, and Harvard University.


\bibliography{wvr_spie}   
\bibliographystyle{spiebib}   

\end{document}